# Unusual nanostructures of "lattice matched" InP on AlInAs

A. Gocalinska[1,*] M.Manganaro,[1] G. Juska,[1] V. Dimastrodonato,[1] K. Thomas,[1] B. A. Joyce,[2] J. Zhang,[2] D. D. Vvedensky,[2] and E. Pelucchi[1]

[1] Tyndall National Institute, University College Cork, "Lee Maltings," Dyke Parade, Cork, Ireland

[2] The Blackett Laboratory, Imperial College London, London SW7 2AZ, United Kingdom

## Abstract

We show that the morphology of the initial monolayers of InP on $Al_{0.48}In_{0.52}As$ grown by metalorganic vapor-phase epitaxy does not follow the expected layer-by-layer growth mode of lattice-matched systems, but instead develops a number of low-dimensional structures, e.g. quantum dots and wires. We discuss how the macroscopically strain-free heteroepitaxy might be strongly affected by local phase separation/alloying-induced strain and that the preferred aggregation of adatom species on the substrate surface and reduced wettability of InP on AlInAs surfaces might be the cause of the unusual (step) organization and morphology.

*Corresponding author

The formation of interfaces with structural, compositional, and morphological integrity is crucial for the performance of many devices. Imperfect interfaces produce broadening in photoluminescence line widths and degrade electronic transport by enhanced scattering. Structure and morphology can be optimized through controlled sample preparation and a judicious choice of growth conditions. However, while semiconductor alloys enable band gaps to be engineered, the attainment of compositional uniformity presents altogether different challenges. Indeed, in III-V systems, phase separation is common when alloys are deposited onto a lattice-matched substrate, for example, by molecular-beam epitaxy (MBE).[1] Our focus here is $Al_{1-x}In_xAs$, a large band-gap (lattice-matched) material used in heterostructures with InP. When produced by MBE (in specific, but a relatively large range of growth conditions), this alloy is known to exhibit clustering when deposited onto InP.[2,3] Interestingly, theoretical studies[4,5] have shown that this type of incipient spinodal decomposition is forbidden if the surface of the alloy film is perfectly flat because of the regions of additional strain created with respect to the random alloy, which has zero mean strain everywhere. But on a surface with roughness, phase separation can become more active at roughness-induced steps due to the accommodation of the additional strain.[4,5,6]

There have also been studies of clustering in III-V systems grown by metalorganic vapor-phase epitaxy (MOVPE)[7,8,9], but none has addressed the systematics of how compositional fluctuations are affected by misorientation or growth conditions. Yet there are reasons to expect that compositional variations in an MOVPE environment may be different from MBE. The (often, but not always observed) high surface mobility of the polyatomic precursors used in MOVPE enables these species to arrive at the steps of even a nominally singular surface, where decomposition occurs preferentially.[10,11,12] On non-planar substrates, the orientation-dependence of the decomposition rate is the origin of growth-rate anisotropies.[13] On InP surfaces, this scenario produces growth by step flow or step bunching over a wide range of growth conditions.[14] As noted above, this can favor phase separation (see also Ref. 15), which, in principle, could affect only the last monolayer.

In this letter, we report the morphology of InP films grown on a macroscopically lattice-matched substrate during MOVPE. We present evidence that epitaxial self-assembled three-dimensional InP islands, which we hereafter refer to as quantum dots (QDs), form on $Al_{0.48}In_{0.52}As$ under what would otherwise be "normal" epitaxial growth conditions. This is an unexpected observation, as these two materials are often grown in complex, lattice-matched, device structures, and their compatibility is limited only by the extent of clustering. Nevertheless, as will become clear in the following, the specific surface organization (and possible phase separation) of $Al_{0.48}In_{0.52}As$ has profound effects on the nucleation of InP (mono)layers, and seems to act as a primary source of the unexpected organization (and/or aggregation). Our results have several important consequences. First, they may explain the difficulties in growing InP/AlInAs multiquantum wells and help to improve the quality of bulk devices containing that interface. Moreover, this opens a new applications window for creating strain-free type II QD structures. For example, type-II heterostructures are attractive systems for both microelectronics and optoelectronics, as the staggered band gap makes the interface energetically favorable for converting photogenerated excitons into free charge carriers. The effect can be amplified and tuned by quantum confinement, so type-II quantum wells[16] and QDs are of substantial interest for possible applications in optical memories (and for quantum information)[17,18], detectors,[19] and solar cells[20]. Finally, our results invite several fundamental questions about the physics of MOVPE and epitaxial growth processes, especially the relation between local atomic arrangements and macroscopic growth morphology. We stress that extensive studies we carried out (see Figure 4 and accompanying text) to ascertain that no unintentional causes (e.g. bad growth conditions inducing defected growth) are responsible of our observation, which are, to our best knowledge, fully induced by "real" fundamental physical processes.

The samples used in this study were grown by MOVPE at low pressure (80 mbar) in a commercial horizontal reactor with purified $N_2$ as the carrier gas. The precursors were trimethylindium (TMIn), trimethylaluminium (TMAl), arsine ($AsH_3$) and phosphine ($PH_3$). For the study of morphology, thin InP films of various thickness were grown on $Al_{0.48}In_{0.52}As$ 120-nm-thick layers following 100 nm of homoepitaxial buffers on (100) $\pm$ 0.02° InP perfectly oriented, or slightly misoriented, semi-insulating substrates. InP buffer growth conditions were optimized, as in Ref. 14. Growth conditions for AlInAs layer were fixed for all the samples: V/III ratio of 110, growth rate G = 1 $\mu$m/hr, real estimated growth temperature $T_g \approx 600$ °C. The growth conditions for the growth of InP layers were: V/III ratio of 180, G = 0.7 $\mu$m/hr, $T_g$ varied in range of 530-665°C. The difference between the buffer and InP growth conditions is related to laboratory history and the complex device structure in which the effect reported here was observed for the first time. All samples were investigated by atomic force microscopy (AFM) in tapping mode to image the surface morphology. Lattice matching of the $Al_xIn_{1-x}As$ layers was confirmed by X-ray diffraction, showing a composition of x = 48% $\pm$ 0.5%, which is in our reproducibility range. To determine the "actual" composition of the dots, wet chemical etching  was performed using nitric and hydrochloric acids solutions in deionised (DI) water (65% $HNO_3$:DI and 37% HCl:DI in the weight ratio specified in the text). Samples were immersed in etching solution for an amount of time indicated in the text, then rinsed with DI water, and blow-dried with nitrogen. A sonic bath was used to facilitate uniform etching.

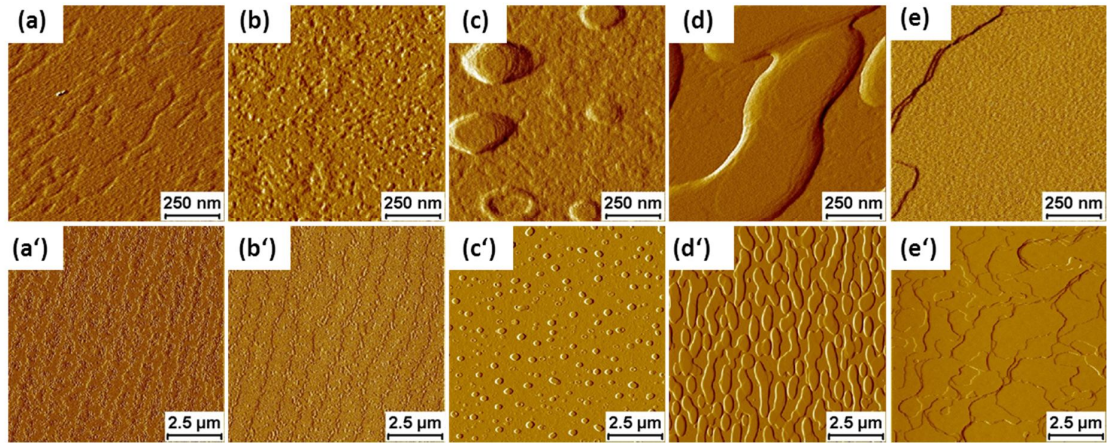

**Figure 1** (Colour online) Surface morphology (AFM signal amplitudes) of samples grown at 630 °C on perfectly oriented substrates with variable InP cap thickness: no InP [(a) and (a')], 0.5 nm [(b) and (b')], 1 nm [(c) and (c')], 4nm [(d) and (d')], 8nm [(e) and (e')].

Figure 1 shows the morphology of the thin InP films grown on lattice-matched AlInAs. These results are far from the epitaxial step-flow expected for perfect lattice matching. For fixed growth conditions, with increasing InP thickness of the InP cap, we observe the formation of what appears to be a thin wetting layer evolving into self-organized dots and ring-like structures, which then coalesce and eventually flatten. The bottom AlInAs layer is always grown under the same conditions for the structures used here; the changes described are only for the cap film. The features have initially grown in all dimensions (for nominal layer thickness up to 1 nm), but the height quickly saturates near 8 nm [Figure 2 (a)], while the features continue to grow laterally. After depositing nominally 8 nm of InP the flat surface fully recovered its normal 2D organization. This unusual morphology was obtained in a broad range of growth temperatures. The QDs appeared in a temperature range from 565 °C to 630 °C [Figure 2 (b)]. There is a discernible trend: at lower growth temperatures, the dot density is much higher, providing greater surface coverage, but with a lower average height. At temperatures below and above the dot formation range, a flat surface was observed, with distinct monolayer islanding (not shown).

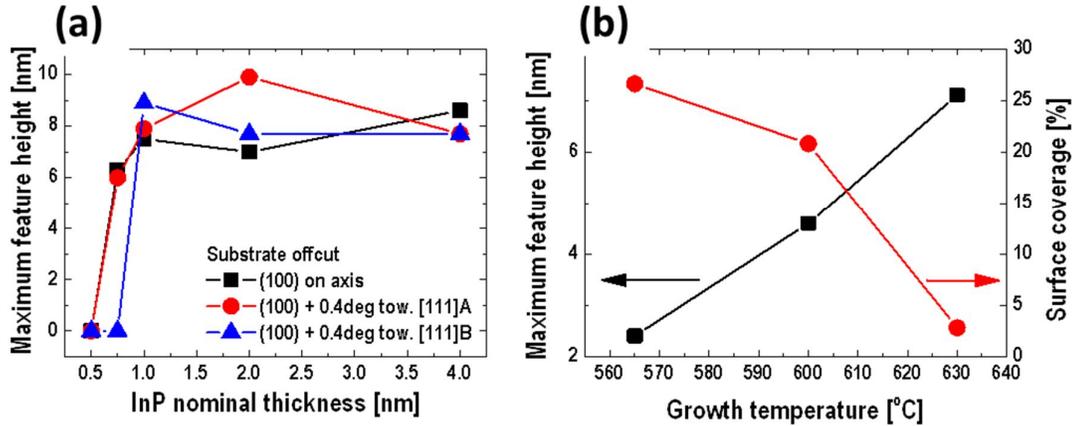

**Figure 2** (Colour online) Influence of growth condition on QDs. (a) Maximum height dependence on the nominal layer thickness grown at 630 °C on wafers with different miscuts). (b) Maximum height and surface coverage variations with the growth temperature for nominally 1nm-thick InP layers on perfectly-oriented substrates.[21]

A small substrate misorientation was found to have a profound impact on the InP surface morphology[14], so we tested how QD formation proceeds on wafers with a small initial off-cut, specifically, 0.4° toward [111]A and [111]B planes. Experiments were carried out in the same range of growth conditions [Figure 2 (a)] with the most striking difference observed at the limits for dot formation (low growth temperature and minimal thickness, 565 °C and 0.75 nm, respectively), where on the perfectly oriented substrate we have observed multiple small dots (of sizes varying over 20-250 nm and aspect ratios from 1 to 10), the growth on 0.4° offcut wafers resulted in either stripes (for B-type surfaces) or a combination of ridges and dots (for A-type surfaces) [Figure 3 (a, b)]. For the lowest thickness, we observed the growth of the QDs on B-type surfaces at 630 °C, which was different from other substrates. While the deposition of nominally ~3 monolayers (MLs) of InP (~0.75 nm) on on-axis and A-type surfaces resulted in QDs, on B-type wafers the morphology remained flat. In other growth conditions, we observed self-organization of the QDs along the surface morphological features on A-type substrates [Figure 3 (c)]. For thicker layers, the step organization of the InP layer does not follow the direction in which the islands merge, while for the A-type surface, the features linked along the step direction, for the B-type surface the process was perpendicular. The full systematics of these phenomena will be presented elsewhere[22].

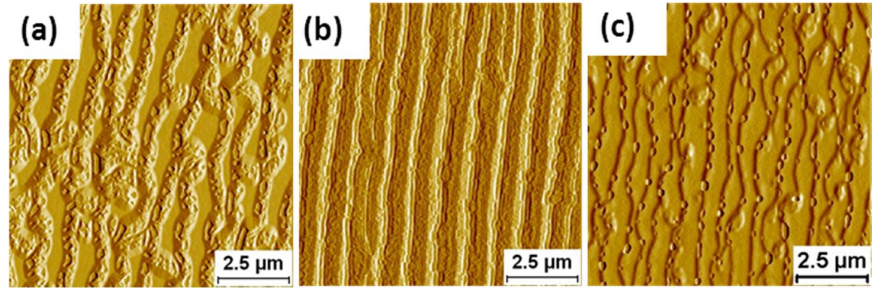

**Figure 3** (Colour online) Surface morphology (AFM signal amplitudes) of samples grown on misoriented substrates of (a) 0.4° toward [111]A, (b) 0.4° toward [111]B (both grown at 565 °C with InP cap thickness of 1 nm), and (c) 0.4° toward [111]A, grown at 600 °C with InP cap thickness of 0.75 nm.

In Figure 4 we present the results of tests conducted to determine how InP wets the underlying AlInAs layer. Firstly, we performed selective wet chemical etching to define the composition of the QDs and confirm the presence of a wetting layer. For a surface uniformly covered with InP, we would expect the nitric acid solution to leave the sample unaffected, since it should etch only alloys containing arsenic. On the other hand, hydrochloric acid, responsive to InP, should remove the cap only and reveal the morphology of the AlInAs beneath. We found that, on samples capped with InP, nitric acid in critical concentrations attacked the material between the dots, etching down to the InP buffer very quickly. Initially this process created pillars crowned with the QDs, and then etched the pillars laterally, allowing the dot to collapse onto the substrate [Figure 4 (a)]. Then it removed the center of each dot, so that just the rings remained on the surface [Figure 4 (b)]. Eventually, all material was removed by prolonged etching and flushing, leaving the step-bunched surface of the underlying epitaxial InP. On the other hand, the HCl etching resulted in a rough, bumped surface, not resembling the uncapped, stepped AlInAs reference [Figure 4 (c)].

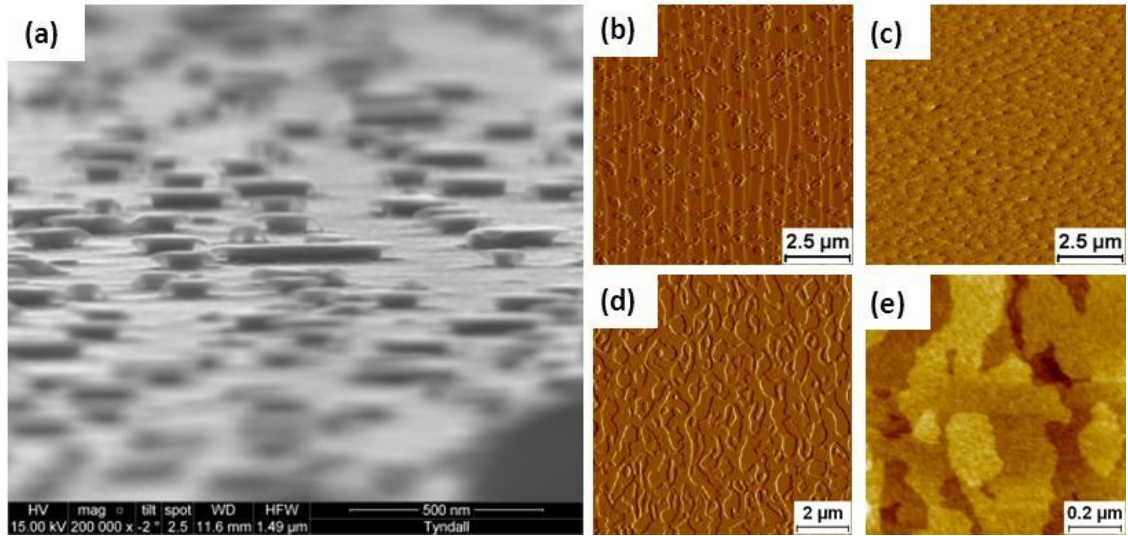

**Figure 4** (Colour online) Sample morphology after wet chemical etching. (a) SEM image of a sample capped with 1nm of InP at 600 °C on an on-axis substrate, etched in 1:1.5 HNO3:DI solution for 15 sec., (b) AFM image (signal amplitude) of a sample capped with 2 nm of InP at 630 °C on substrate misoriented by 0.4° toward [111]B etched in 1:1.5 HNO3:DI solution for 30 sec., (c) AFM image (signal amplitude) of a sample capped with 4 nm of InP at 630 °C on substrate misoriented by 0.4° toward [111]A etched in 1:1 HCl:DI solution for 30 sec.; (d) AFM image (signal amplitude) of 2 nm of InP on 0.25 nm of AlInAs grown at 630 °C on a perfectly oriented substrate, (e) AFM image (signal height) of 2 nm of InP on 2 nm of InGaAs on AlInAs on a perfectly oriented substrate.

The wet chemical etching suggests that the composition of the QDs might not be uniform. Since, during the nitric acid treatment, the dot initially works as an etching mask while the acid removes the center of each dot, this would indicate that the middle-bottom part of each dot disc contains an arsenic alloyed compound, and the remaining rings are formed of nearly pure InP. Consistent with this picture is the fact that the hydrochloric acid leaves the surface covered with small nanometric bumps with a density that corresponds to the initial dot density (the alloyed arsenic containing center is the only feature left). Subsequently, we carried out an additional series of growth experiments with fixed growth conditions, changing the thickness of the AlInAs layer from ~0.5 monolayers up to the bulk (several nanometer thick layers). The results on layers as thin as 2 nm were identical to those previously shown for > 100 nm films. The thinner layers, however, show significant spatial segregation of overgrowth in that InP seems to grow preferentially in some parts of the sample, that is, not wetting the AlInAs film uniformly. Even 1 ML [Figure 4 (d)] is sufficient, to provide the conditions for nonuniform growth. However, introducing even a thin (<2nm) lattice-match InGaAs layer between AlInAs and InP resulted in flat, uniform growth [Figure 4 (e)] We see this (and other not shown experiments) as an additional proof that the original AlInAs surface is a normal quality surface, and the observed segregation of InP is not induced by artificial causes, like unintentional defects.

We have no straightforward explanation for the observed behavior. The system seems to lack any overall significant strain: high-resolution X-ray diffraction measurements (not shown, and in absence of a microscopic transmission electron microscopy analysis, which might give more detailed insight) confirm that the AlInAs layer is macroscopically fully strained and that there has been no significant relaxation by dislocation formation or by any other means. Furthermore, it is tetragonally distorted and the layer is constrained in the z-direction, i.e. the in-plane lattice parameter is that of InP (a more comprehensive analysis will be presented elsewhere[22]). Experiments with very thin AlInAs layers exclude the possibility of introducing significant strain, as a single monolayer would not be expected to be anything but pseudomorphic and the otherwise standard/correct growth conditions (these layers when grown, separately, homo- or heteroepitaxially on $In_{0.53}Ga_{0.47}As$ are perfectly uniform) should result in smooth epitaxial layers. Also the growth with a lattice matched InGaAs insert contributes to that conclusion, as the resulting strain would not be much changed by its addition, while the formation of InP nanostructures is not observed anymore.

On the other hand, surface reconstruction mechanisms in InP (001) are known to differ from all other III–V semiconductor materials as the structural transformations do not necessarily follow the trend considered otherwise universal for III–V semiconductors[23,24]. The lower bonding energy between two In atoms in respect to In-P bond might lead to interfacial clustering of In, possibly contributing to non-planar growth in the first several monolayers.[25]

We must also consider the fact that, even though the $Al_{0.48}In_{0.52}As$ alloy yields an average lattice constant equal to that of InP, this is a spatial average with no information about the distribution of Al and In. There are two extreme cases: complete mixing, and complete phase separation. Both would have the same average lattice constant, but would be very different substrates in terms of the strain distribution presented to the next InP layers. The reality is most likely somewhere in between, with some clustering, so there are regions where local strain is large, and regions where it is small. There is also the issue of atomic size. Indium is a larger atom than aluminum, so even in the case of perfect mixing, the surface would appear corrugated. All somehow coherent with the fact that we observe different surface organization when differently miscut substrates (i.e. steps and corrugation) are chosen.

Furthermore, while interfacial alloying (and possibly, induced local strain and alloy segregation or a Stranski–Krastanov-like process) seems to have a role in initiating the nucleation of InP dot-like structures, which are somehow favored by the reduced wettability of InP on AlInAs surfaces during MOVPE, effectively retarding 2D nucleation, the constant lateral growth of the ring/dot structures seems to be linked to significant adatom migration, preferentially aggregating (or nucleating) around the pre-existing islands. Comprehensive microscopy work will be needed to assess the exact role of all these variables and will be the subject of future investigations, and all this will be likely to require an extensive theoretical analysis to be fully understood.

In conclusion, we have shown, that the epitaxy of the initial several monolayers of InP on $Al_{0.48}In_{0.52}As$ does not follow the expected layer-by-layer ordered growth mode. While our finding point to a role of local alloying-induced strain, on the other hand the formation of nearly strain-free, lattice matched QDs and wires (at least macroscopically, in the sense of not having effects on the subsequent layers) seems to have been obtained, and we reported the preferred aggregation of adatom species on the substrate surface as one of the contributing mechanisms.

As we discussed in our introduction, InP/AlInAs heterostructures are good candidate materials for optoelectronic devices, Schottky barrier technology and CMOS implementations. Moreover, the lattice matching of $Al_{0.48}In_{0.52}As$ to InP does not put a limit on the structure thickness and promises easy stacking of multiple QWs/QDs without any elastic strain. Finally, despite the technological interest, there is scarce evidence in the literature of successful growth of multiple InP/AlInAs QWs by MOVPE.[26] The results presented here could be an indication of a reason for this.

### Acknowledgments

This research was enabled by the Irish Higher Education Authority Program for Research in Third Level Institutions (2007-2011) via the INSPIRE programme, by Science Foundation Ireland under grants 10/IN.1/I3000, 07/SRC/I1173 and by an EU project FP7-ICT under grant 258033 (MODE-GAP).